\documentstyle[aaspp4,epsf]{article}
\newcommand{\figcomment}[1]{#1}
\def\unsetyr{\def\oyear{\relax}\def\cyear{\relax}\def\cyeara{a\relax}\def\cyearb{b\relax}}
\def\setyr{\def\oyear{(}\def\cyear{)}\def\cyeara{a)}\def\cyearb{b)}}
\unsetyr
\def\jcite#1{\setyr\cite{#1}\unsetyr}
\def\rmmat#1{{\hbox{\rm #1}}}

\newcommand{\be}{\begin{equation}}
\newcommand{\ee}{\end{equation}}
\newcommand{\ba}{\begin{eqnarray}}
\newcommand{\ea}{\end{eqnarray}}
\newcommand{\ie}{{\it i.e.~}}
\newcommand{\eg}{{\it e.g.~}}
\newcommand{\comment}[1]{\relax}
\def\eqref#1{Equation~\ref{eq:#1}}
\def\figref#1{Figure~\ref{fig:#1}}

\begin{document}
\newcommand{\bfi}{{\bf B}} \newcommand{\efi}{{\bf E}}
\newcommand{\lel}{{\lambda_e^{\!\!\!\!-}}}
\newcommand{\me}{m_e}
\newcommand{\mcs}{{m_e c^2}}
\title{Do magnetars glitch? : Timing irregularities in anomalous X-ray pulsars}
\author{Jeremy S. Heyl}
\authoremail{jsheyl@tapir.caltech.edu}
\affil{Theoretical Astrophysics, mail code 130-33,
California Institute of Technology, Pasadena, CA 91125}
\author{Lars Hernquist}
\authoremail{lars@ucolick.org}
\affil{Lick Observatory,
University of California, Santa Cruz, California 95064, USA}

\begin{abstract}
We examine the timing history of several anomalous X-ray pulsars
(AXPs), and find that they exhibit spin down irregularities
statistically similar to those of radio pulsars.  We propose that
these irregularities are simply glitches like those of radio pulsars
and fit glitching spin down solutions to the data available for
1E~2259+586, 1E~1048.1-5937 and 4U~0142+61.  The inferred magnitude of
the glitches ($\Delta \Omega$) is comparable to that exhibited in radio
pulsar glitching.  With these results we argue that the three AXPs
that we studied are isolated neutron stars spinning down by magnetic
dipole radiation and powered by a combination of cooling, magnetic
field decay and magnetospheric particle bombardment.  \end{abstract}


\section{Introduction}

Over the past decade, several X-ray pulsars with unusually long
periods ($P \sim 10$~s) and unusually small luminosities ($L \sim
10^{35}-10^{36}$~erg/s) have been discovered in or near young
supernova remnants (\eg \cite{Vasi97b}).  \jcite{Mere95} grouped this
objects as a class.  These objects are known as braking
(\cite{Whit96}) or ``anomalous'' (\cite{vanP95}) X-ray pulsars
(AXPs). \jcite{Thom96} suggest that these objects are isolated neutron
stars spinning down by magnetic dipole radiation with $B_p \gtrsim
10^{14}$~G, \ie ``magnetars'' (\cite{Dunc92}) and that the decay of
their intense magnetic fields powers their X-ray emission.
\jcite{Heyl98decay} examine this issue further in the context of the
cooling evolution of these objects.  \jcite{Heyl97kes} and
\jcite{Heyl98rxj} argue that their emission may be powered by the
cooling of the neutron-star core through a strongly magnetized or
light-element envelope without any appreciable field decay.  The
observations by \jcite{Kouv98} of SGR 1806-20 indicate that SGRs and
AXPs may be related, which further connects the AXPs with models of
SGRs that invoke magnetars (\cite{Thom95}; \cite{Thom96}).  For
several AXPs, the spin-down age of the neutron star and the age of the
surrounding remnant are similar, bolstering these arguments.

Several contrasting models for AXPs have been proposed in which a
neutron star is slowly accreting from the ISM (\cite{Wang97}), from a
very low mass companion (\cite{Mere95}; \cite{Bayk96}) or from the
remains of a high-mass X-ray binary (\cite{vanP95}; \cite{Whit96};
\cite{Ghos97} ).  In these models, irregularities in the accretion
flow explain the irregular nature of the spin down or torque noise
(\cite{Bayk96}) as found in observations of accreting pulsars (\eg
\cite{Bild97}; \cite{Nels97}; \cite{Chak97}).  However, it is
difficult to reconcile the ages of the supernova remnants with the
slow rotation rates of the pulsars, unless these neutron stars were
born rotating unusually slowly.

In this {\it Letter}, motivated by the coincidences between the
characteristic spin-down ages of some AXPs and the ages of the SNRs
surrounding them, we examine a model in which AXPs are isolated
pulsars rather than accreting sources.  Specifically, we will focus on
the spin-down irregularities of three anomalous X-ray pulsars and
compare their timing behavior with that of isolated radio pulsars
having a range of periods from 1.6~ms to 3.7~s.

\section{Timing Irregularities in AXPs and Radio Pulsars}

\jcite{Arzo94} study the timing behavior of large sample of radio
pulsars over a wide range of periods and period derivatives.  To
characterize the timing irregularities of these objects, they fit the
observed rotational phase of each pulsar with a function of the form 
\be
\phi = \phi_0 + \nu t + \frac{1}{2} {\dot \nu} t^2 + \frac{1}{6}
{\ddot \nu} t^3 \ldots .
\label{eq:phase}
\ee
Further, they define a stability parameter
\be
\Delta(t) = \log \left ( \frac{1}{6\nu} |{\ddot \nu}| t^3 \right ),
\ee
if $|{\ddot \nu}| > 2 \sigma_{\ddot \nu}$; otherwise they adopt an
upper limit 
\be
\Delta(t) < \log \left ( \frac{2 \sigma_{\ddot \nu} t^3}{6 \nu}
\right ).
\ee
Since the observations were taken over several years, they use
a characteristic time of $10^8$~s so that $\Delta_8 \equiv \Delta(10^8
\rmmat{s})$ is approximately the logarithm of the clock error of the
pulsar in seconds over the span of the observations.

For the AXPs, we use the first derivative of \eqref{phase} to
determine the value of $\Delta_8$ for the pulsars 1E~2259+586
(\cite{Bayk96}), 1E~1048.1-5937 (\cite{Oost98}) and 4U~0142+61
(\cite{Hell94}; \cite{Whit96} and the August 1984 observation analyzed
by \cite{Isra94}). \jcite{Isra94} present several marginal results for
other observations of 4U~0142+61 which we omit from our analysis.

\begin{figure}
\figcomment{\plottwo{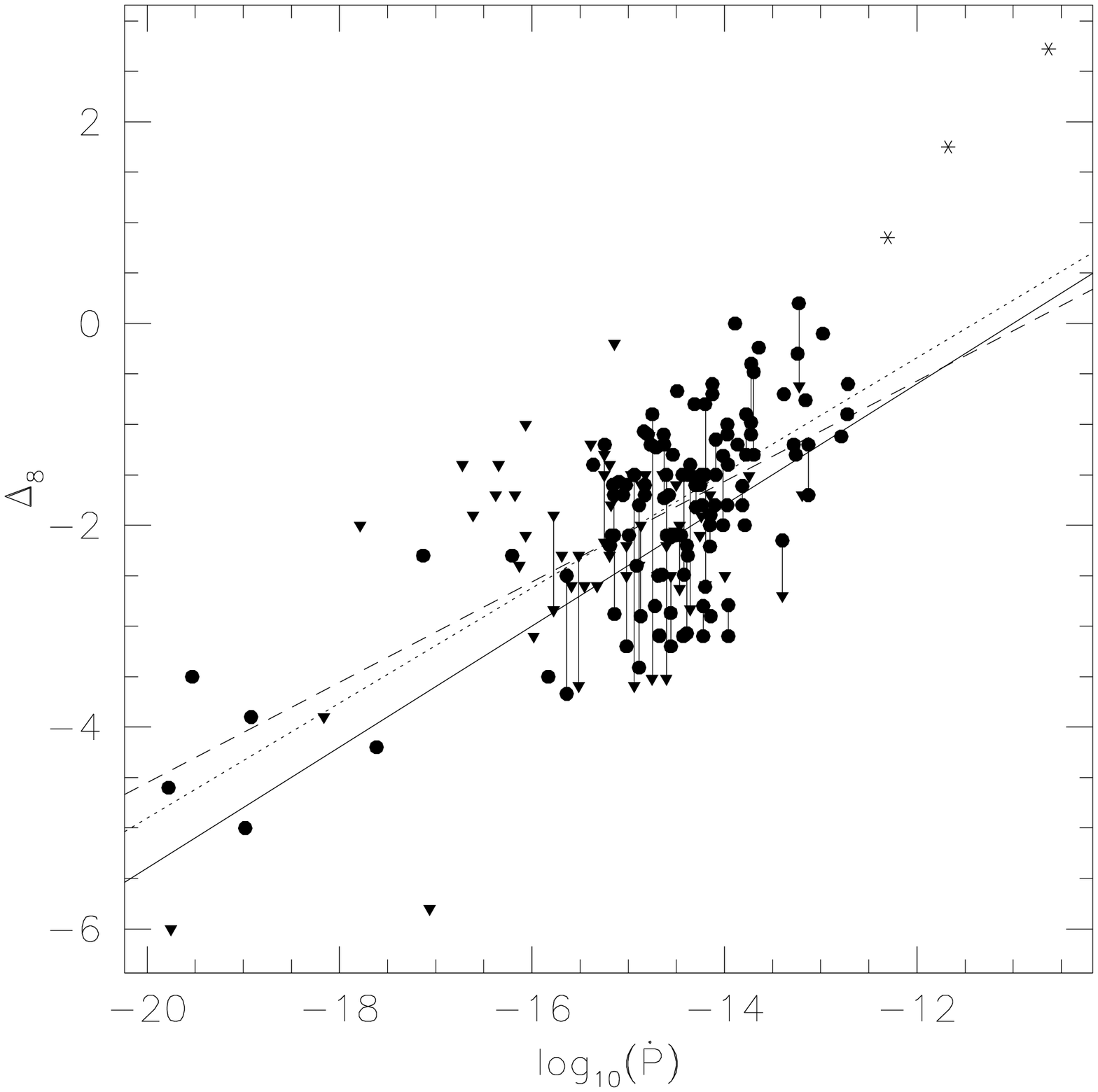}{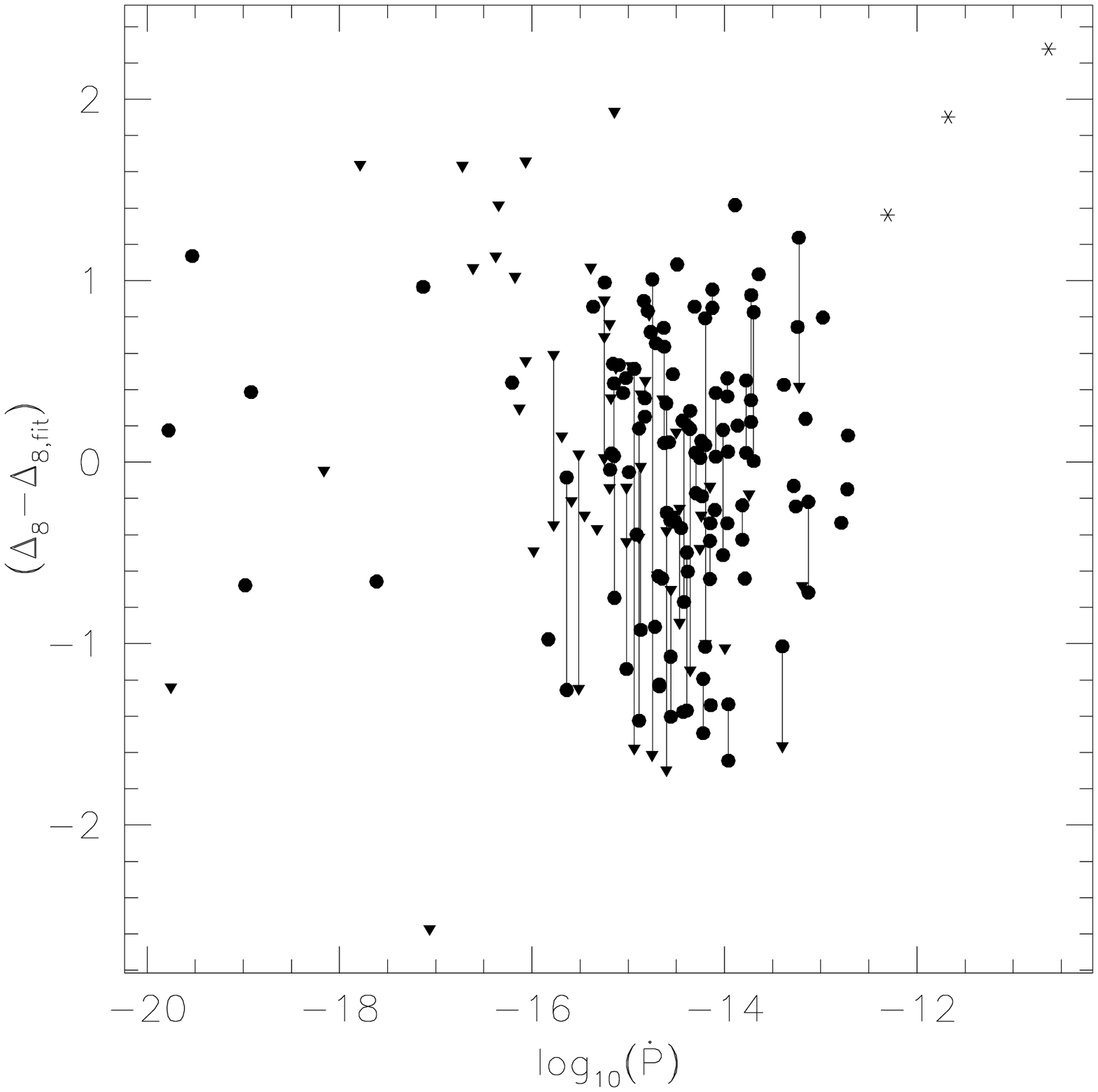}}
\caption{The left panel depicts the timing noise parameter,
$\Delta_8$, plotted against the period derivative for 139 radio
pulsars (triangles and disks from \cite{Arzo94}) and 3 anomalous X-ray
pulsars (asterisks).  Inverted triangles represent upper limits.  When
two or more estimates of $\Delta_8$ are available, the corresponding
points are connected by a vertical line. The right panel shows the
residuals of the data relative to the best-fit linear relation (the
dotted line).}
\label{fig:activity}
\end{figure}

\figref{activity} presents the results for 139 radio pulsars from
\jcite{Arzo94} along with the three AXPs studied here.  In the left
panel, the lower solid line traces the relation $\Delta_8 = 6.6 + 0.6
\log {\dot P}$ given by \jcite{Arzo94} for the set of radio pulsars.
The upper dotted line gives the best fit relation for the complete
sample, $\Delta_8 = 6.5 + 0.57 \log {\dot P}$, and the long-dashed
line traces the relation $\Delta_8 = 5.4 + 0.50 \log {\dot P}$, which
is the best-fit relationship for the radio pulsars alone.

The right panel shows the residuals of the data relative to the
best-fit linear relation for the entire dataset.  Although the AXPs
(asterisks) have significantly larger values of ${\dot P}$ and
activity parameters, they follow the relationship for the pulsar
population as a whole well.

\subsection{Glitch Fitting}

\begin{figure}
\figcomment{
\plottwo{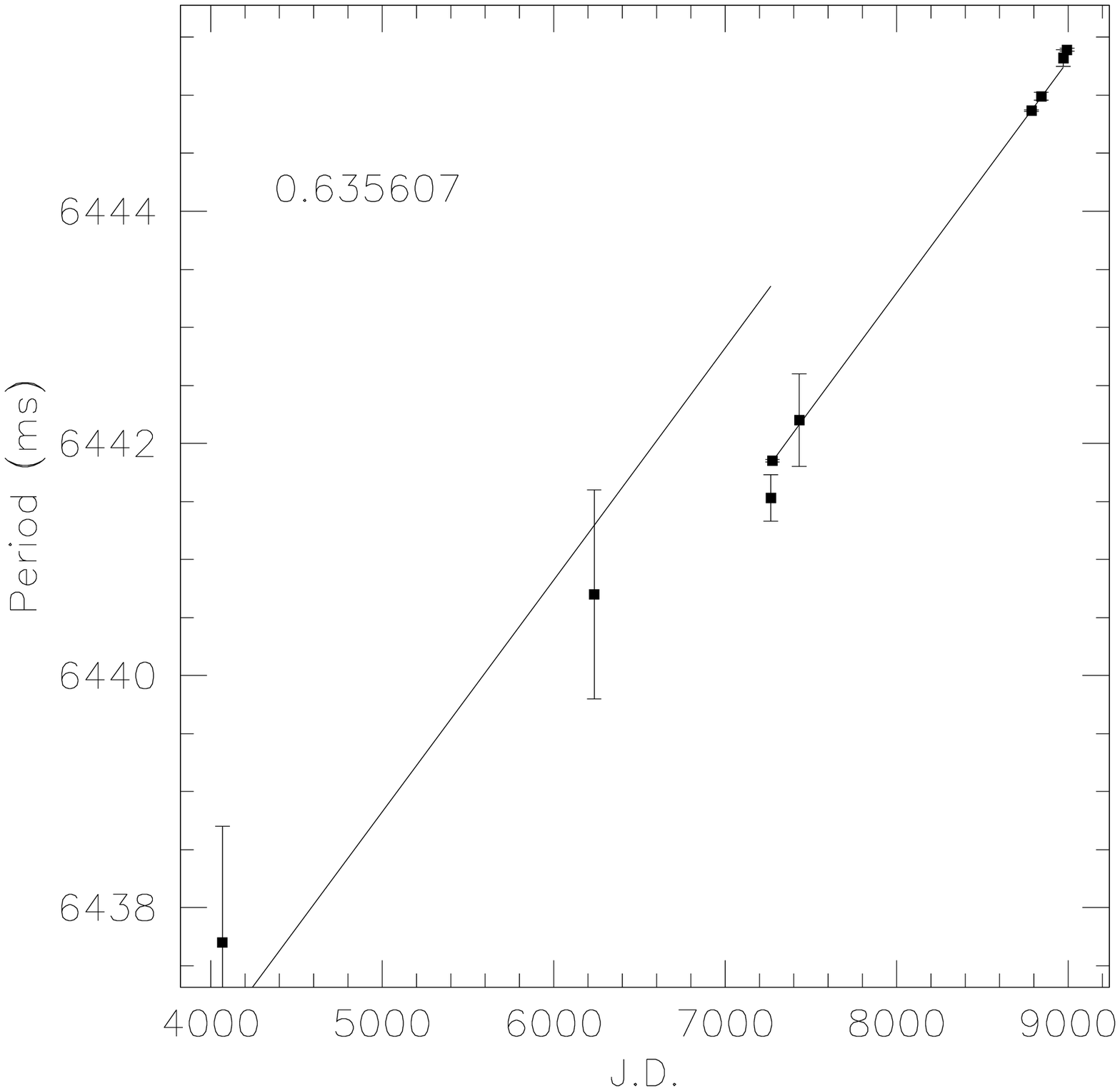}{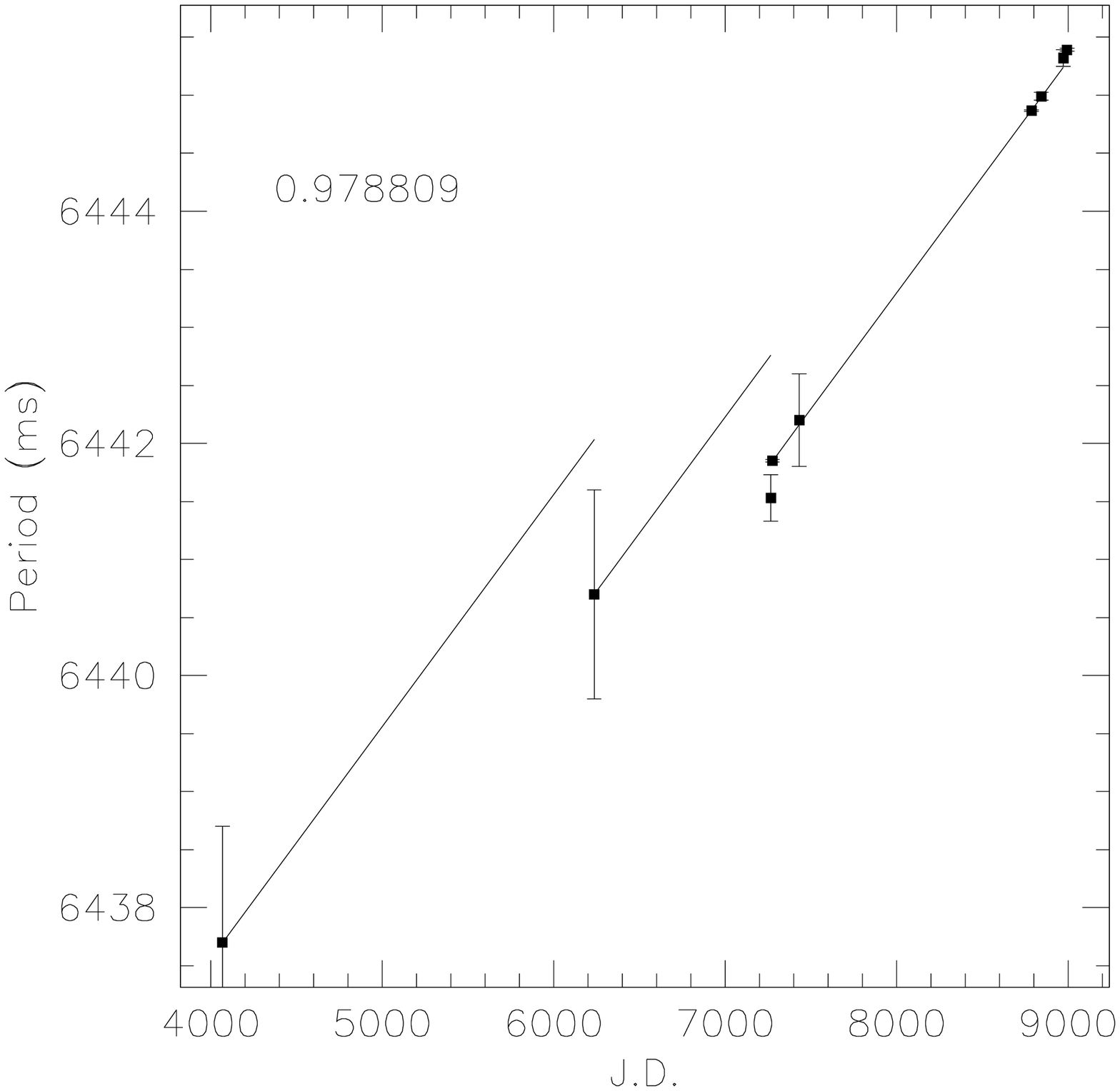}

\plottwo{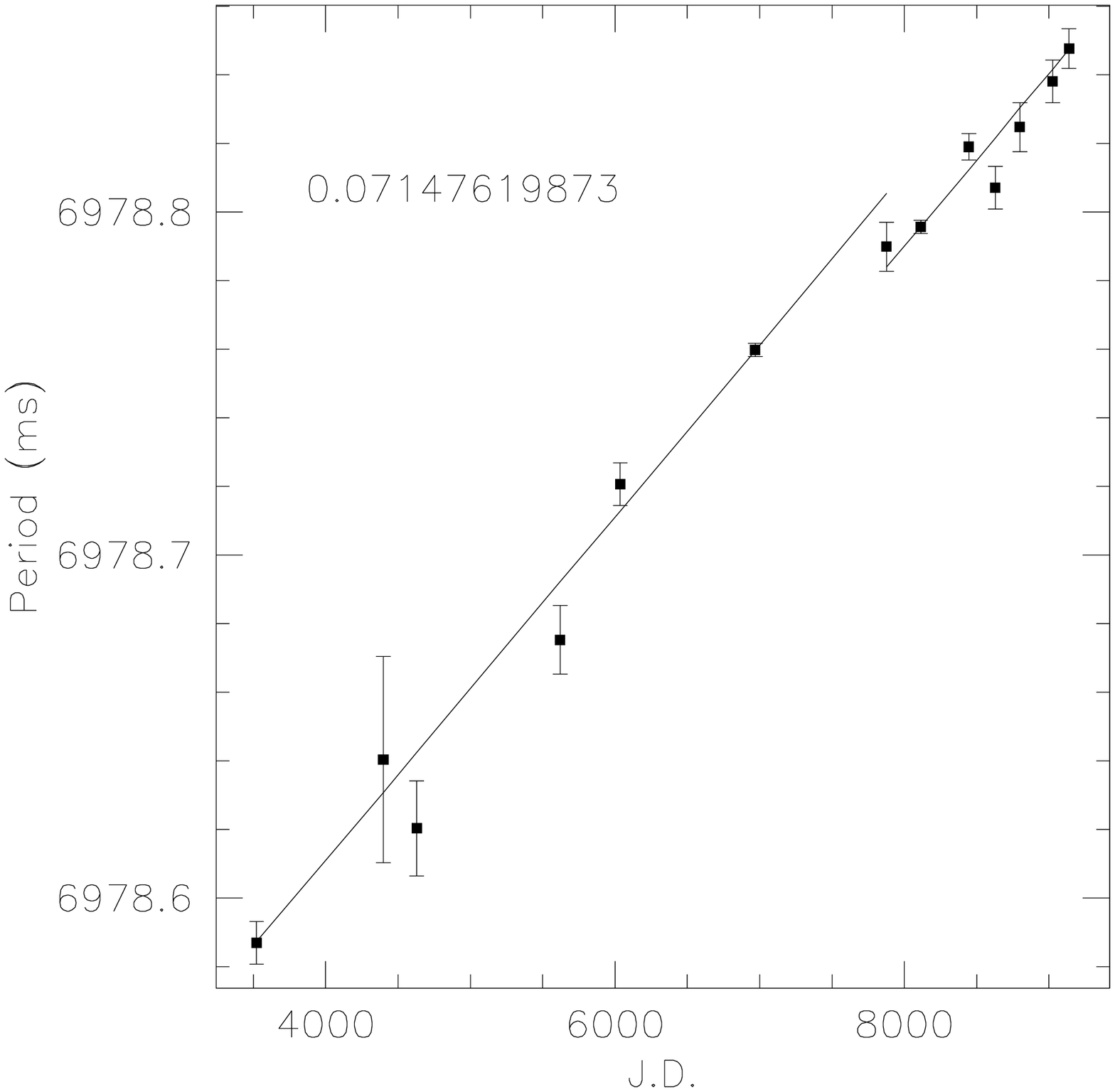}{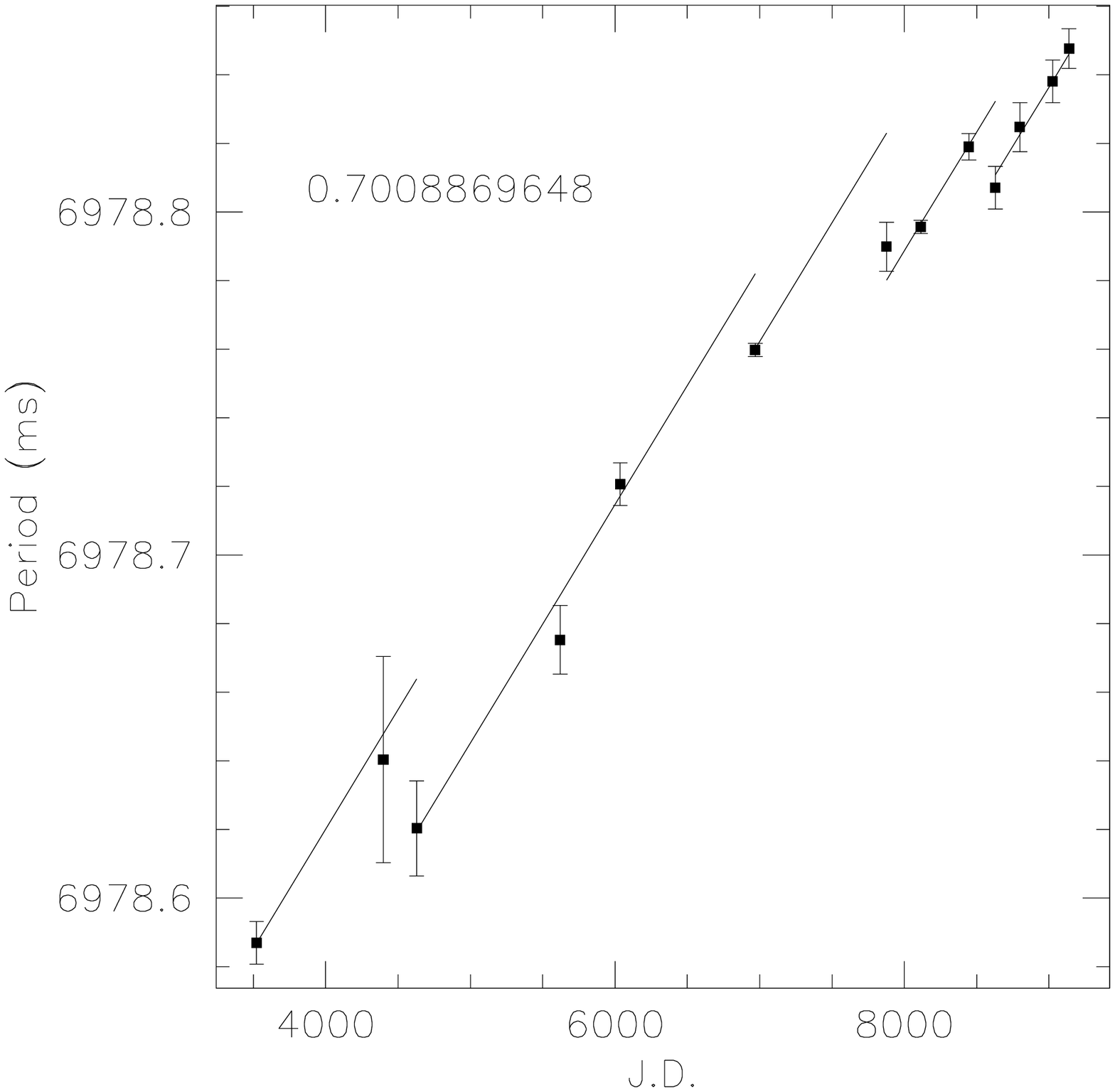}
}
\caption{Glitch models for the AXPs 1E~1048.1-5937 and 1E~2259+586.
The upper two panels depict one- and two-glitch models for the
spin-down of 1E~1048.1-5937, and the lower panels depict one- and
five-glitch models for the period evolution of 1E~2259+586.  The
$\chi^2$ likelihood that each model fits the data is given
in the upper left quadrant of each panel. }
\label{fig:glitch}
\end{figure}

For two of the AXPs there are sufficient observations that we can fit
glitches to the timing solution to determine a lower limit on the
frequency of glitching in these objects and an upper limit on the
magnitude of the glitches themselves.  We fit a spin-down relationship
linear in time with the possibility of several discontinuous jumps in
period between the observations.  Since the data are sparsely sampled,
we assume that the period can jump immediately before any observation.
The best fit glitch models are found by minimizing the value of
$\chi^2$ with respect to the slope and intercepts of the model while
varying the positions of the discontinuities.  \figref{glitch} gives
two glitch models for the AXPs 1E~1048.1-5937 and 1E~2259+586.

Although both pulsars have periods between 6 and 7 seconds, their
period derivatives differ by nearly a factor of thirty.  When the
timing data for these pulsars are fitted with glitch solutions, we
find that the characteristic size of the glitches also varies by a
factor of approximately thirty.  The typical size of the glitches for
1E~1048.1-5937 is $\Delta \Omega \approx 2 \times 10^{-4}$~s$^{-1}$,
while the more slowly decelerating pulsar, 1E~2259+586, glitches more
weakly with $\Delta \Omega \approx 5 \times 10^{-6}$~s$^{-1}$,
comparable to the value found by \jcite{Usov94}.  The
strengths of these glitches are comparable to the glitches in the Vela
and Crab pulsars respectively (\eg \cite{Shap83}).

\section{Discussion}

We have found that AXPs follow an extension of the relation between
pulsar clock error and period derivative proposed by \jcite{Arzo94}
for millisecond and ordinary radio pulsars.  Furthermore, by fitting
discontinuities on a linear relationship, $P(t)$, we find that AXPs
exhibit glitches with values of $\Delta \Omega$ similar to radio
pulsars.  For example, if glitches occur when vortex lines pinned to
nuclei in the crust are freed and angular momentum is suddenly
transferred to the crust (\eg \cite{Alpa84}; \cite{Pine85}), $\Delta
\Omega$ is simply proportional to the number of vortex lines which
become unpinned during the event.  In this scenario, the value of
$\Delta \Omega$ may depend on the properties of the outer core and
inner crust of the neutron star and will not be strongly affected by the
presence of a strong magnetic field.  \jcite{Thom96} argue that the
evolution of magnetic field in a strongly magnetized neutron star may
encourage glitching with an amplitude $\Delta \Omega \sim 10^{-5}$,
similar to the values inferred here.

In this {\it Letter}, we have described similarities between the AXPs
and ordinary radio pulsars in the context of timing irregularities.
This begs the question of why AXPs are not observed as radio pulsars.
A simple possibility is that the solid angle subtended by the
open-field lines in a pulsar magnetosphere decreases as the period
increases.  Both models and observations of the radio pulsars indicate
that the radio emission is limited to the open-field lines;
consequently, it becomes less likely that the pulse beam crosses our
line of sight as the period of the pulsar increases.

\jcite{Mesz92} summarizes several models for the emission beams from
radio pulsars in the context of observations and theory.  In general,
the half-opening angle of the pulse beam is $\sim 5^\circ$ for a
period of 1~s and decreases as $P^{-\gamma}$ where $\gamma = 1/3 - 1/2.$
To be conservative we will compare two cone-type pulsars for which
\be
\rho \sim 6.5^{\circ} \left ( \frac{P}{1~\rmmat{s}} \right )^{-1/3}.
\ee

The probability to observe a given pulsar will be proportional to the
solid angle subtended by its rotating beam or beams,
\begin{equation}
{\cal P} \sim \sin \rho
\end{equation}
For a seven-second pulsar, ${\cal P} \sim$ 6\%; 
consequently, if the AXPs do emit radio waves like cone pulsars,
about dozen of these objects would need to be found before we
should expect to have a fifty-fifty chance of detecting radio emission
from any one of them. 

In preceding paragraphs, we argued why it might be unlikely to observe
the radio emission from AXPs.  Most pulsar emission models rely on the
formation of an electron-positron cascade through one-photon pair
production to power the radio emission (\eg \cite{Mesz92}).  Several
studies have argued that if $B \gtrsim B_c$ where $B_c \approx 4.414
\times 10^{13}$~G, a gamma-ray photon will preferentially decay into a
bound electron-positron pair (\cite{Usov96}).  This creates a
relativistically outflowing neutral gas which is unable to produce
radio emission collectively (\cite{Aron98}).  As the gas flows
into more weakly magnetized regions, the positronium becomes more
weakly bound and eventually photoionizes.  Depending on where the pair
plasma forms, the outflow may power a plerion.  The AXPs may therefore
be connected with the plerions which appear to be driven by a strong
field but lack a radiative signature of a central compact source
(\cite{Helf95}).  Because of the formation of bound $e^+e^-$ pairs in
strong fields, it is natural that AXPs with $B>B_c$ are unobserved in
the radio; however, like other young isolated pulsars, they may
exhibit plerionic emission.

Alternatively, \jcite{Bari97b} argue that in sufficiently strong fields
($B \gtrsim B_c$) the QED process of photon splitting (\cite{Adle71};
\cite{Heyl97hesplit}) can dominate one-photon pair production.  This
will effectively quench the pair cascade, making coherent pulsed radio
emission impossible.  

\section{Conclusion}

Because the AXPs studied in this paper exhibit spin down variations
like neutron stars which we know to be isolated, our results suggest
that they are isolated as well.  This implies that these objects do
indeed have $B \sim 10^{14-15}$~G and that their X-ray emission is
powered either by the cooling of the core of the neutron star or by
the decay of its field, since these are the two largest sources of
free energy available to these objects.  The possibility of
distinguishing between this possibility and alternative models in
which variations in AXP periods are driven primarily by accretion from
a disk may soon be possible when detailed spectra of AXPs are obtained
with the AXAF satellite.

\acknowledgements

We would like to thank Jon Arons and S. Kulkarni for valuable
discussions, and Zaven Arzoumanian for kindly providing the
\jcite{Arzo94} results in machine readable form along with a macro to
reproduce the figure presented in their paper.  J.S.H. acknowledges a
Lee A. DuBridge postdoctoral scholarship.


\cleardoublepage

\figcomment{
\end{document}
\end
}

\cleardoublepage

Figure 1 Left 
\plotone{activity.eps}
\cleardoublepage
Figure 1 Right
\plotone{activit2.eps}
\cleardoublepage
Figure 2 Upper Left
\plotone{1e1048_3.eps}
\cleardoublepage
Figure 2 Upper Right
\plotone{1e1048_5.eps}
\cleardoublepage
Figure 2 Lower Left
\plotone{1e2259_2.eps}
\cleardoublepage
Figure 2 Lower Right
\plotone{1e2259_5.eps}
\end{document}